\documentstyle[12pt,aaspp4,flushrt]{article}

\slugcomment{Submitted to ApJ}

\def\beq{\begin{equation}}
\def\eeq{\end{equation}}
\def\bey{\begin{eqnarray}}
\def\eey{\end{eqnarray}}
\def\beqarray{\begin{eqnarray}}
\def\eeqarray{\end{eqnarray}}

\def\phix{\phi_x}
\def\phiy{\phi_y}
\def\Dd{D_{\rm d}}
\def\Ds{D_{\rm s}}
\def\Dds{D_{\rm ds}}
\def\zd{z_{\rm d}}
\def\H0{H_0}
\def\gammac{\gamma_{1}}
\def\gammas{\gamma_{2}}

\def\mpc{\,{\rm {Mpc}}}
\def\kms{\,{\rm {km\, s^{-1}}}}

\begin{document}

\title{Analytic Time Delays and $H_0$ Estimates \\ for Gravitational Lenses}
\author{Hans J.~Witt$^{1}$, Shude Mao$^{2}$, and Charles R.~Keeton$^{3}$}
\affil{$^{1}$ Astrophysikalisches Institut Potsdam, An der Sternwarte 16,
  14482 Potsdam, Germany}
\affil{$^{2}$ University of Manchester, Jodrell Bank Observatory,
  Macclesfield, Cheshire SK11 9DL, UK}
\affil{$^{3}$ Steward Observatory, University of Arizona, Tucson, AZ 85721, USA}

\begin{abstract}
We study gravitational lens time delays for a general family
of lensing potentials, which includes the popular singular
isothermal elliptical potential and singular isothermal
elliptical density distribution but allows general angular
structure.  Using a novel approach, we show that the time
delay can be cast in a very simple form, depending only on
the observed image positions.  Including an external shear
changes the time delay proportional to the shear strength,
and varying the radial profile of the potential changes the
time delay approximately linearly.  These analytic results
can be used to obtain simple estimates of the time delay and
the Hubble constant in observed gravitational lenses.  The
naive estimates for four of five time delay lenses show
surprising agreement with each other and with local
measurements of $\H0$; the complicated Q~0957+561 system
is the only outlier.  The agreement suggests that it is
reasonable to use simple isothermal lens models to infer
$\H0$, although it is still important to check this
conclusion by examining detailed models and by measuring
more lensing time delays.
\end{abstract}


\section {Introduction}

Refsdal (1964) first proposed that time delays between images in
multiply-imaged gravitational lenses can be used to measure the
Hubble constant $\H0$.  This method is attractive because it is a
single-step process and is based on the well-established theory
of General Relativity.  After a long ordeal, this method is finally
beginning to bear fruit.  In the last few years, the time delays
in six gravitational lenses have been measured, which yield
$\H0 \approx 65\pm 15 \kms\mpc^{-1}$ (e.g., Koopmans \& Fassnacht
1999; Browne 2000).  The lensing measurement is an important
way of confirming and extending local determinations of $\H0$
(see Freedman 1999 for a review), which are still subject to
systematic uncertainties such as the LMC distance, metallicity
effects, and photometric contamination (e.g., Stanek, Zaritsky
\& Harris 1998; Kennicutt et al.\ 1998; Mochejska et al.\ 1999).
At present, the error budget in the lensing measurement is
dominated by systematic uncertainties in the lens modeling (see
Keeton et al.\ 2000a for a discussion).  Ultimately the accuracy
may be limited by the uncertainties induced by the large-scale
structure along the line of sight at the few percent level
(Seljak 1994; Barkana 1996; Schneider 1997), although if
random these effects should shrink as $N^{-1/2}$ as the number
$N$ of lenses with measured time delays increases.

Lensing measurements of $\H0$ are typically derived from models
based on isothermal galaxies, because such models are consistent
with individual lenses, lens statistics, stellar dynamics, and
X-ray galaxies (e.g., Fabbiano 1989; Maoz \& Rix 1993; Kochanek
1995, 1996; Grogin \& Narayan 1996; Rix et al.\ 1997).  Modelers
usually adopt a parametrized form, either an isothermal elliptical
potential (e.g., Blandford \& Kochanek 1987) or an isothermal
elliptical density (e.g., Kassiola \& Kovner 1993; Kormann,
Schneider \& Bartelmann 1994; Keeton \& Kochanek 1998), and
adjust the parameters to fit the data.\footnote{However, see
Saha \& Williams (1997) and Blandford, Surpi \& Kundi\'c (2000)
for novel non-parametric methods.}  The purpose of this paper
is to gain insights into the time delays in a more general
family of lens models that includes these commonly used
isothermal models and their variants.

The outline of this paper is as follows.  In section 2, we show
that the time delays for singular isothermal elliptical potential
(SIEP) and singular isothermal elliptical density (SIED)
distributions have a remarkably simple and elegant form, and that
the result actually holds for a general family of potentials with
the form $\phi = r {\cal F}(\theta)$.  In section 3, we show how
the time delay is affected when we incorporate an external shear
and change the radial profile of the potential.  In section 4,
we combine our analytic results with data for the time delay
lenses to estimate $\H0$.  Finally in section 5, we offer a
summary and discussion.

\section{Time Delay for Generalized Isothermal Models}

The time delay in gravitational lenses is given by (e.g., Schneider,
Ehlers \& Falco 1992)
\beq \label{eq:time}
\Delta t (x,y) = \frac{D}{2c} (1+\zd) \left[ (x-\xi)^2 + (y-\eta)^2
- 2 \phi(x,y) \right], ~~D \equiv {\Dd \Ds \over \Dds},
\eeq
where $\zd$ is the redshift of the lens, $(\xi, \eta)$ is the angular
source position, $(x, y)$ is the angular image position, $\Dd$ and
$\Ds$ are angular diameter distances from the observer to the lens
and source, respectively, and $\Dds$ is the angular diameter
distance from the lens to the source. The dimensionless potential
$\phi$ satisfies the two-dimensional Poisson equation 
\beq
\nabla^2\phi(x,y) = 2 \kappa, \quad
\kappa = 
{\Sigma(x, y) \over \Sigma_{\rm cr}}\ , \label{eq:sigcr}
\eeq
where $\Sigma$ is the projected surface mass distribution of the
lens and $\Sigma_{\rm cr} = {c^2 \Ds / (4 \pi G \Dd \Dds)}$ is the
critical surface density for lensing.

To proceed further we adopt a specific potential or density form.
Individual lenses and lens statistics are usually consistent
with isothermal models (e.g., Maoz \& Rix 1993; Kochanek 1995,
1996; Grogin \& Narayan 1996), so it is common to adopt either
the isothermal elliptical potential,
\beq \label{eq:sep}
\quad \phi = \frac{a_0}{2} (x^2+y^2/q^2)^{1/2},
\eeq
or the elliptical density distribution,
\beq \label{eq:sed}
\nabla^2 \phi = 2 \kappa, \quad
\kappa = \frac{a_0}{2q} (x^2+y^2/q^2)^{-1/2}, 
\eeq
where $a_0$ specifies the overall angular size of the lens.
For the SIED, the lens potential is given by, e.g., Kassiola
\& Kovner (1993), Kormann, Schneider \& Bartelmann (1994),
and Keeton \& Kochanek (1998).  Note that we have chosen a
coordinate system that is centered on the lens galaxy and
aligned the $x$-axis along the lens galaxy's major axis.  For
simplicity, we have also assumed that the isothermal potential
and density distributions are singular at the origin, although
we return to this issue briefly in \S 5.

In this paper, we study a more general family of lens models.
The potential is assumed to obey the relation
\beq \label{eq:phi}
\phi = x \phix + y \phiy,
\eeq
where $\phix$ and $\phiy$ are the first derivatives of the
potential, which are just the components of the deflection
angle in the $x$ and $y$ directions.  In polar coordinates
$(r, \theta)$, eq.~(\ref{eq:phi}) can be written in the
simple form
\beq \label{eq:phiPolar}
\phi = r {\partial \phi \over \partial r},
\eeq
whose general solution is 
\beq \label{eq:phiSolution}
\phi = r {\cal F}(\theta),
\eeq
where ${\cal F}(\theta)$ is an arbitrary function of $\theta$.
This potential corresponds to a density of the form
\beq \label{eq:isokappa}
\kappa = \frac{1}{2r} {\cal G}(\theta), \quad
{\cal G}(\theta) = {\cal F}(\theta) + {\partial^2{\cal F}\over\partial\theta^2} .
\eeq
Thus eq.~(\ref{eq:phi}) describes a family of scale-free models
that includes both the SIEP and SIED models but allows for general
angular structure; in other words, these are generalized isothermal
models.  We now show that this family is extremely useful in
deriving the time delays between images.

The lens equation is given by
\beq \label{eq:lens} 
\xi  = x - \phix, \quad
\eta = y - \phiy.
\eeq
Substituting eqs.~(\ref{eq:phi}) and (\ref{eq:lens}) into
the time delay expression in eq.~(\ref{eq:time}), we obtain
\beq
\Delta t (x,y) = \frac{D}{2c} (1+z_d) \left[ (x-\xi)^2 + (y-\eta)^2 \\
- 2 x (x-\xi) - 2 y (y-\eta) \right],
\eeq
which can be further simplified into
\beq
\Delta t (x,y) = \frac{D}{2c} (1+z_d)\,
(\xi^2 + \eta^2 - x^2 -y^2).
\eeq
Since only the relative time delay is observable, we compute
the time delay between two images 
$i$ and $j$, which is given by
\beq \label{sedDelay}
\Delta t_{i,j} = \frac{D}{2c} (1+z_d) (r_j^2 - r_i^2),
\eeq
where $r_i = (x_i^2 + y_i^2)^{1/2}$ is simply the distance of
image $i$ from the center of the galaxy.  This surprisingly
simple expression is valid for all lens potentials satisfying
eq.~(\ref{eq:phi}), including both the SIEP and SIED models
as well as more general angular structures.\footnote{After the
completion of our paper, we learned that eq.~(\ref{sedDelay})
has been derived for particular cases by Koopmans, de Bruyn \&
Jackson (1998) and Zhao \& Pronk (2000) using more complicated
methods.}  Since $r_i^2$ and $r_j^2$ are rotationally invariant,
eq.~(\ref{sedDelay}) is valid even after an arbitrary rotation.
In other words, provided we know the center of the lens galaxy,
the predicted time delay can be computed using eq.~(\ref{sedDelay})
irrespective of the lens orientation and without any need to
search for the best fit model parameters.  Since the predicted
time delay scales as $\propto \H0^{-1}$, it can be compared with
a measured time delay to determine $\H0$; we demonstrate this
method in \S 4.
 
For completeness we note that the following relations hold for
a potential which satisfies eq.~(\ref{eq:phi}):
\beq
x \phi_{xx} + y \phi_{xy} = 0,~
x \phi_{xy} + y \phi_{yy} = 0,~
\phi_{xx} \phi_{yy} - \phi_{xy}^2 = 0.
\eeq
The last expression can be used to simplify the magnification
of the images:
\beq
\mu^{-1} = (1-\phi_{xx}) (1-\phi_{yy}) - \phi_{xy}^2
= 1- \phi_{xx} -\phi_{yy} = 1 - 2 \kappa,
\eeq
revealing a simple relation between the magnification and the 
surface density in this class of potentials.
  
\section{Time Delay in Other Potentials}

\subsection{Time delay in the presence of shear}

In many gravitational lenses the images cannot be fit without
the inclusion of a tidal perturbation from objects near the
lens galaxy or along the line of sight (e.g., Keeton, Kochanek
\& Seljak 1997; Witt \& Mao 1997).  To lowest order, the
perturbation can be modeled as an external shear with potential
\beq \label{shearPhi}
\phi_\gamma = -{1 \over 2} \gamma r^2 \cos 2(\theta-\theta_\gamma)
= -{1 \over 2} \left[ \gammac (x^2-y^2) + 2 \gammas x y \right],
\eeq
where $\gamma$ is the strength of the shear and $\theta_\gamma$
is its direction with respect to the lens galaxy's major axis,
while
$\gammac=\gamma\cos 2\theta_\gamma$
and 
$\gammas=\gamma\sin 2\theta_\gamma$.
The total potential is then $\phi_{\rm tot}=\phi+\phi_\gamma$,
and the lens equation can be written as
\beq \label{leqshear}
\xi  = x - \phix + \gammac x + \gammas y, \quad
\eta = y - \phiy + \gammas x - \gammac y.
\eeq
Following the same logic as in \S 2, we obtain the time delay
\beq
\Delta t (x,y) = \frac{D}{2c} (1+z_d) \left[ \xi^2 + \eta^2 
- r^2 - r^2 \gamma \cos 2(\theta - \theta_\gamma) \right].
\eeq
The relative time delay between two images $i$ and $j$ in the
presence of shear is then
\beq \label{eq:dtshear}
\Delta t_{i,j} = \frac{D}{2c} (1+z_d)
\left\{
(r_j^2 - r_i^2) +
\gamma \left[
   r_j^2 \cos 2(\theta_j-\theta_\gamma)
 - r_i^2 \cos 2(\theta_i-\theta_\gamma)
\right]
\right\}.
\eeq
This time delay depends on the shear amplitude and direction
and therefore cannot be determined without detailed modeling.
Nevertheless, we can make several remarks.  The change in the
time delay (relative to the no-shear case) is proportional to
$\gamma$.  For two-image lenses that have images at different
distances ($r_i \ne r_j$) and a small shear, the shear should
have a small effect on the time delay.  However, when the images
lie at approximately the same distance ($r_i \approx r_j$),
such as in some four-image lenses, the shear term may be
significant.  In particular, perturbation theory reveals that
in the presence of a shear the distance of an image from the
critical curve scales $\propto\gamma$ (to lowest order; e.g.,
Kochanek 1991), so the two terms in eq.~(\ref{eq:dtshear}) can
be comparable.

Although we cannot determine the time delay without knowing
the shear amplitude and direction, we can put bounds on it.
If the angle between the shear and the lens galaxy major axis
is allowed to take on any value, then the time delay is
bounded by
\beq \label{eq:Tbounds1}
  T_{i,j}^{(-)} \le \Delta t_{i,j} \le T_{i,j}^{(+)} ,
\eeq
where
\beq \label{eq:Tbounds2}
  T_{i,j}^{(\pm)} = \frac{D}{2c} (1+z_d)
\left\{
(r_j^2 - r_i^2) \pm
\gamma \left[
   r_i^4 + r_j^4 - 2 r_i^2 r_j^2 \cos 2(\theta_i-\theta_j)
\right]^{1/2}
\right\}.
\eeq
If the images are directly opposite each other ($|\theta_i -
\theta_j| = 180^\circ$, as in a circular lens), the bounds are
$T_{i,j}^{(\pm)} = (1\pm\gamma) \Delta t_{i,j}^0$ where
$\Delta t_{i,j}^0$ is the no-shear time delay from
eq.~(\ref{sedDelay}).  This means that, for example, a 10\%
shear ($\gamma=0.1$) leads to a 10\% uncertainty in the time
delay (and hence $H_0$) if there is no information about the
angle between the shear and the lens galaxy major axis.  We
apply eqs.~(\ref{eq:Tbounds1}) and (\ref{eq:Tbounds2}) to obtain
bounds for specific lenses in \S 4.

\subsection{Time delay in non-isothermal models}

Observed lenses seem to be consistent with isothermal galaxies
(see introduction), but lenses with images at similar distances
from the lens galaxy can often be modeled with other density
profiles as well.  Models of PG~1115+080 and B~1608+656 show
that steeper density profiles lead to larger predicted time
delays and hence larger values for $\H0$ (Keeton \& Kochanek
1997; Impey et al.\ 1998; Koopmans \& Fassnacht 1999).  We
analyze deviations from an isothermal profile by relaxing the
condition in eq.~(\ref{eq:phi}) to
\beq \label{eq:phibeta}
\beta \phi = x \phix + y \phiy,
\eeq
where $\beta$ is a constant.  (Isothermal models have $\beta=1$.)
This condition can be understood by finding the general solution
for the potential,
\beq \label{eq:phiBeta}
\phi = r^\beta \cal{F}(\theta),
\eeq
which corresponds to a density of the form
\beq
\kappa = \frac{1}{2} r^{\beta-2} {\cal G}(\theta), \quad
{\cal G}(\theta) = \beta^2 {\cal F}(\theta) +
  {\partial^2{\cal F}\over\partial\theta^2} .
\eeq
So this model has a radial power law for the potential and
density, and $\beta$ is the power law slope of the potential.
Relaxing eq.~(\ref{eq:sep}) to fit eq.~(\ref{eq:phibeta}), an
obvious (but not unique) family of solutions is given by
$\phi \propto (x^n + y^n/q^n)^{\alpha/2}$, with $\beta =
n \alpha/2 = {\rm const}$. For $n>2$, the iso-potential
contours become boxier, while for $n <2$, the contours become 
diskier. These contour shapes mimic those seen in the inner
parts of elliptical galaxies (e.g., Bender et al.\ 1989).

More generally, for any potential satisfying eq.~(\ref{eq:phibeta})
the time delay is given by
\beq
\Delta t (x,y) = \frac{D}{2c} (1+z_d) \left[ (\xi^2 + \eta^2)
+ 2\frac{(1 - \beta)}{\beta} (\xi x + \eta y)
- \frac{(2-\beta)}{\beta} (x^2 +y^2) \right].
\eeq
The time delay between two images $i$ and $j$ is then
\beq \label{eq:dtbeta1}
\Delta t_{i,j} = \frac{D}{2c} (1+z_d) \left\{
 \frac{2-\beta}{\beta}\left(r_j^2-r_i^2\right)
 +2\frac{(1-\beta)}{\beta}\Bigl[\xi(x_i-x_j)+\eta(y_i-y_j)\Bigr]
\right\}.
\eeq
Alternatively, if we substitute for the source position
$(\xi,\eta)$ using the lens equation we find
\beq \label{eq:dtbeta2}
\Delta t_{i,j} = \frac{D}{2c} (1+z_d) \left[ (r_j^2-r_i^2)
 + 2 (1-\beta) (\phi_j-\phi_i) \right],
\eeq
where $\phi_i=\phi(r_i,\theta_i)$.  Eqs.~(\ref{eq:dtbeta1})
and (\ref{eq:dtbeta2}) show that when the model is not
isothermal ($\beta \ne 1$) we cannot eliminate the need
for modeling to determine the source position and/or the
potential at each image.

We computed time delays for non-isothermal elliptical
potentials numerically, and we found that for small to
moderate ellipticities the delays are well approximated by
the isothermal time delay (eq.~\ref{sedDelay}) modified by
a multiplicative factor.  For opposed images
($|\theta_i-\theta_j| \sim 180^\circ$), the time delay is
approximately
$\Delta t_{i,j} \approx [2-\beta]\, \Delta t_{i,j}^{\rm iso}$,
while for orthogonal images
($|\theta_i-\theta_j| \sim 90^\circ$), the time delay is
approximately
$\Delta t_{i,j} \approx [(2-\beta)/\beta]\, \Delta t_{i,j}^{\rm iso}$.
Similar scalings were found by Refsdal \& Surdej (1994) and
Witt, Mao \& Schechter (1995, eq.~8).  By contrast, the
isothermal time delay is not a good approximation for close
image pairs ($|\theta_i-\theta_j| \sim 0$), because the images
are necessarily near the lensing critical curve and hence more
sensitive to the particular lens model.

\section{Implications for $H_0$}

Lensing time delays are interesting because they offer a
measurement of the Hubble constant independent of the distance
ladder.  Unfortunately, the method has proven to be somewhat
more problematic than expected.  Converting a time delay to
$\H0$ requires a lens model, and it is often difficult to
find a model that is both a good fit to the data and well
constrained.  At present we are in the paradoxical situation
that better observational data seem to {\it increase\/} the
uncertainties in $\H0$, because they require lens models that
are more complex and thus harder to constrain (see, e.g.,
Bernstein \& Fischer 1999; Keeton et al.\ 2000a).

As a result, some have advocated applying a simple and physically
motivated class of models (such as isothermal ellipsoids) to
an ensemble of time delay lenses and using the scatter in $\H0$
estimates to evaluate whether the result is robust (e.g.,
Koopmans \& Fassnacht 1999; Browne 2000; E.~Turner, private
communication).  The crucial assumption is that simple but
statistically poor fits may be close enough to the truth to give
reliable $\H0$ estimates -- or if not, the discrepancy will show
up as a large scatter in $\H0$ estimates.  This approach yields
a tantalizing concordance of lensing estimates at $\H0 \approx
65\pm15 \kms\mpc^{-1}$ (Koopmans \& Fassnacht 1999; Browne 2000),
a value that is consistent with values from other methods.  However,
this value is based on a compilation of models from the literature
that are very heterogeneous and include numerous explicit and
implicit biases from the choice and quality of the data and the
choice of classes of lens models.

Our analytic time delays now offer the ability to estimate $\H0$
for the time delay lenses with a method that is not only extremely
simple but also uniform, i.e.\ derived using a uniform set of
modeling assumptions and using the same type of data (image
positions) for all lenses.  We use the data for five of the time
delay lenses (summarized in Table 1) to compare the observed and
predicted time delays and determine $\H0$.\footnote{We exclude
B~1608+656 because its two lens galaxies imply a potential that
does not obey eq.~(\ref{eq:phi}); see Koopmans \& Fassnacht (1999).}
Figure 1 shows the $\H0$ estimates for the generalized isothermal
models discussed in \S 2, assuming no external shear.
Eq.~(\ref{sedDelay}) makes it essentially trivial to compute
$\H0$ from easily measured quantities without any modeling.
Moreover, the estimates depend only on the assumption of an
isothermal profile for the lensing potential (and not on its
angular structure).  Finally, the $\H0$ estimates depend only
weakly on the adopted cosmology. 

With the exception of Q~0957+561, the naive $\H0$ estimates are
consistent with each other and with values from other methods.
Q~0957+561 stands out because the lens is complicated: the lens
galaxy is embedded in a $\sigma \sim 700\kms$ cluster and
appears not to have a simple isothermal profile, two features
that invalidate the class of potentials assumed in
eq.~(\ref{eq:phi}) (see Bernstein \& Fischer 1999; Keeton et
al.\ 2000a; and references therein).  Still, the agreement among
the remaining four systems is surprising given that we have
done no modeling.  On the one hand, the agreement is somewhat
misleading because two of the lenses have large systematic
uncertainties in the lens galaxy position that lead to enormous
$\H0$ errorbars (B~0218+357 and PKS~1830$-$211, which are
discussed in detail by Leh\'ar et al.\ 1999).  On the other
hand, the agreement between PG~1115+080 and B~1600+434 is
intriguing because one has an elliptical lens galaxy and the
other has an edge-on spiral lens galaxy.  Moreover, the rough
agreement between the $\H0$ estimates from the two time delays
in PG~1115+080 provides a useful consistency check on the model
assumptions.

Figure 1 suggests three conclusions about how to strengthen
lensing's ability to provide an independent $\H0$ estimate.
First, the dependence of the analytic time delays on the image
positions emphasizes the importance of precise astrometry of
not only the lensed images but also the lens galaxy.  At
present, the large $\H0$ uncertainties in B~0218+357 and
PKS~1830$-$211 are dominated by the 60--80 mas uncertainties
in the lens galaxy positions.  Second, the agreement between
the naive $\H0$ estimates for four of the five systems
suggests that the lens galaxies have a common density profile;
furthermore, the agreement of the lens results with results
from other methods suggests that assuming the generalized
isothermal model is not unreasonable.  (The same conclusion
was reached by Koopmans \& Fassnacht 1999.)  A good way to
test these conclusions is to increase the number of lenses
with measured time delays and see whether the naive $\H0$
estimates continue to agree.

Third, there are still systematic uncertainties that can only
be addressed with detailed modeling.  Such modeling is clearly
required for Q~0957+561.  More generally, the modeling is
required to fully understand the systematic uncertainties in
the models and their effect on $\H0$.  Models of PG~1115+080
reveal the importance of a group of galaxies surrounding the
lens galaxy, whose effects cannot be modeled as a simple
external shear; the models yield $\H0 = 44\pm4 \kms\mpc^{-1}$
(assuming an SIED galaxy; Impey et al.\ 1998), which is
somewhat smaller than the naive $\H0$ estimate due in part
to the gravitational focusing (or convergence) provided by
the group.  Although increasingly detailed models may be
subject to degeneracies, the recent detection of the host
galaxies of several lensed quasars (see Rix et al.\ 2000 for
a summary) offers many new constraints that should break the
model degeneracies and yield robust lens models (see Keeton
et al.\ 2000ab; Blandford et al.\ 2000).  Even if we can
never break the degeneracies in all time delay lenses,
comparing detailed models with our naive analytic $\H0$
estimates in a few systems will test whether the estimates
are close enough to the truth to be useful.

One important systematic effect not included in Figure 1 is
external shear, but our analytic results even permit an
estimate of its effects (using eqs.~\ref{eq:Tbounds1} and
\ref{eq:Tbounds2}).  Figure 2 shows the bounds on $\H0$ for
various values of the shear strength $\gamma$ for the five
lenses, assuming no knowledge of the angle between the shear
and the lens galaxy major axis.  The bounds would be narrower
given constraints on this angle.  An important qualitative
result is that the $\H0$ estimate from the four-image lens
PG~1115+080 is much more strongly affected by shear than
those from two-image lenses, as expected from the fact that
in four-image lenses the images tend to lie at approximately
the same distance from the lens galaxy (see \S 3.1).

\section{Conclusions}

We have shown that for generalized isothermal models (with
a potential of the form $\phi = r {\cal F}(\theta)$), the
time delay between images can be expressed in a common and
surprisingly simple form involving only the observed image
positions (eq.~\ref{sedDelay}).  An external shear changes
the time delay by an amount proportional to the shear strength,
and affects 4-image lenses more than 2-image lenses.  Changing
the radial profile of the lens galaxy changes the time delay
approximately by a multiplicative factor.  In this paper, we
have only considered singular potential and density
distributions.  This is justified because so far no convincing
faint central image has been observed, and this sets stringent
limits on the central core radius (Wallington \& Narayan 1993;
Kochanek 1996).  Numerically we find that such small core
radius changes the time delays negligibly ($\lesssim 1\%$).

Our simple expression for the time delay can be combined with
directly observable quantities to give an estimate for the
Hubble constant $\H0$ without the need for any modeling.  The
analytic $\H0$ estimates for four of five time delay lenses
agree surprisingly well with each other and with distance
ladder measurements of $\H0$.  The outlier, Q~0957+561, is
known to be a complicated system where the lensing potential
is inconsistent with our assumptions of a pure isothermal
model.  Thus it appears that simple isothermal lens models
are often reasonable for obtaining lensing estimates of $\H0$
(see also Koopmans \& Fassnacht 1999).

Still, it is crucial to test this conclusion more rigorously,
and there are three ways to do so.  First, the astrometry for
two of the time delay lenses (B~0218+357 and PKS~1830$-$211)
must be significantly improved to reduce the uncertainties
and verify the $\H0$ concordance.  Second, more time delays
should be measured (and good astrometric precision obtained
for those lenses) to test whether all of the $\H0$ estimates
derived from isothermal lens models are mutually consistent.
Increasing the sample size will also reduce the random
uncertainties due to large-scale structure.  Third, more
individual lenses should undergo detailed modeling to check
the $\H0$ estimates obtained from isothermal models and
understand the systematic uncertainties.  These detailed
models are most instructive when they are constrained by
observations of the host galaxies of the lenses quasars,
because requiring models to reproduce the distortion of the
host galaxy significantly improves the ability to find a
robust and well-constrained model (see Keeton et al.\ 2000ab;
Blandford et al.\ 2000).

Finally, we recall that the dependence of the time delay
on the density profile of the lens galaxy affects the $\H0$
estimates (also see Keeton \& Kochanek 1997; Koopmans \&
Fassnacht 1999).  If lens galaxies tend to have profiles
steeper than isothermal ($\phi(r) \propto r^\beta$ with
$\beta<1$), then by insisting on using isothermal models
we tend to underestimate $\H0$ (or vice versa).  There are
two possible situations:
\begin{itemize}
\item
{\it Galaxies have a common density profile (little or no
scatter in $\beta$).\/}  If so, the $\H0$ estimates obtained
by applying isothermal models to numerous lenses would show
little scatter.  If these estimates were to agree with distance
ladder measurements of $\H0$, it would suggest that galaxies
generally have isothermal profiles.
\item
{\it Galaxies have a scatter in $\beta$.\/}  If so, isothermal
lens $\H0$ estimates would show a significant scatter.  It is
possible that lenses are biased tracers of $\beta$: on the
one hand, lenses with larger values of $\beta$ tend to produce
higher magnifications and hence have a larger magnification
bias, while on the other hand they have smaller cross sections
for lensing (e.g., Kochanek 1991; Wambsganss \& Paczy\'nski
1994; Witt et al.\ 1995; Witt \& Mao 2000).  However, when
lenses are normalized to reproduce the image separations in
observed lenses, these two effects essentially cancel (see
Kochanek 1996).  Thus it appears that lenses are not strongly
biased with regard to $\beta$, so the scatter in lensing $\H0$
estimates would be similar to the scatter in $\beta$.
\end{itemize}
The present data suggest the first situation, namely that
lenses are generally consistent with isothermal galaxies.
By collecting more and better data, however, we will be
able to test this conclusion more rigorously and obtain
a robust and independent measurement of $\H0$.

\acknowledgements
We thank Ian Browne and Peter Wilkinson for stimulating
discussions, and Chris Kochanek, Peter Schneider, and Ed Turner
for helpful comments on the manuscript.

%
%

\clearpage

\centerline{\epsfxsize=6.0in \epsfbox{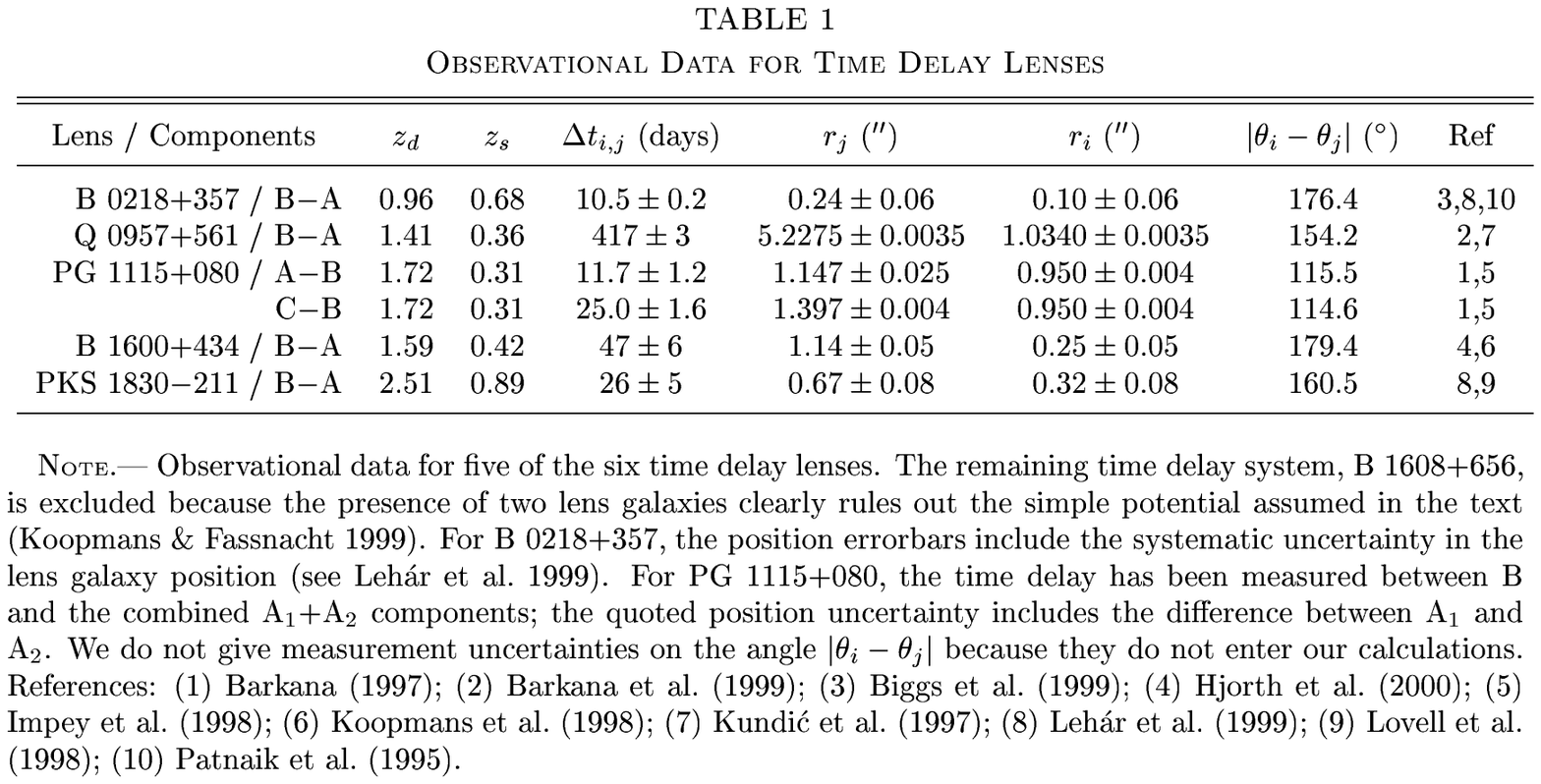}}

\begin{figure}[h]
\centerline{\epsfxsize=5.5in \epsfbox{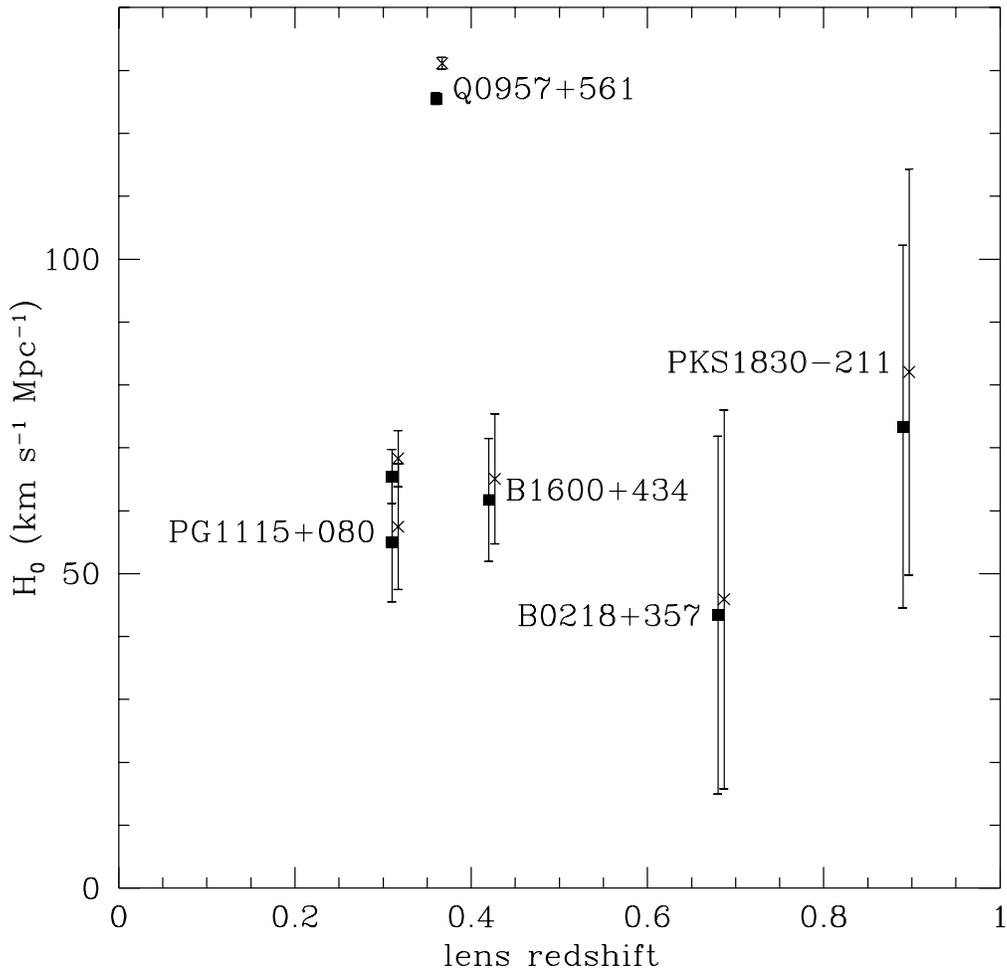}}
\caption{
Naive $\H0$ estimates for the five time delay lenses listed in
Table 1.  The estimates are computed using eq.~(\ref{sedDelay}),
i.e.\ under the assumption that the lensing potential obeys
eq.~(\ref{eq:phi}) and there is no external shear.  The errorbars
indicate uncertainties due to measurement errors in the time
delays and image positions.  The filled squares are computed
assuming a cosmology with $\Omega_M=1$ and $\Omega_\Lambda=0$,
while the crosses (offset slightly in redshift) are computed
assuming $\Omega_M=0.3$ and $\Omega_\Lambda=0.7$.  For
PG~1115+080, there are two $\H0$ estimates because there are
two measured time delays among the four images.
}
\end{figure}

\begin{figure}[h]
\centerline{\epsfxsize=5.5in \epsfbox{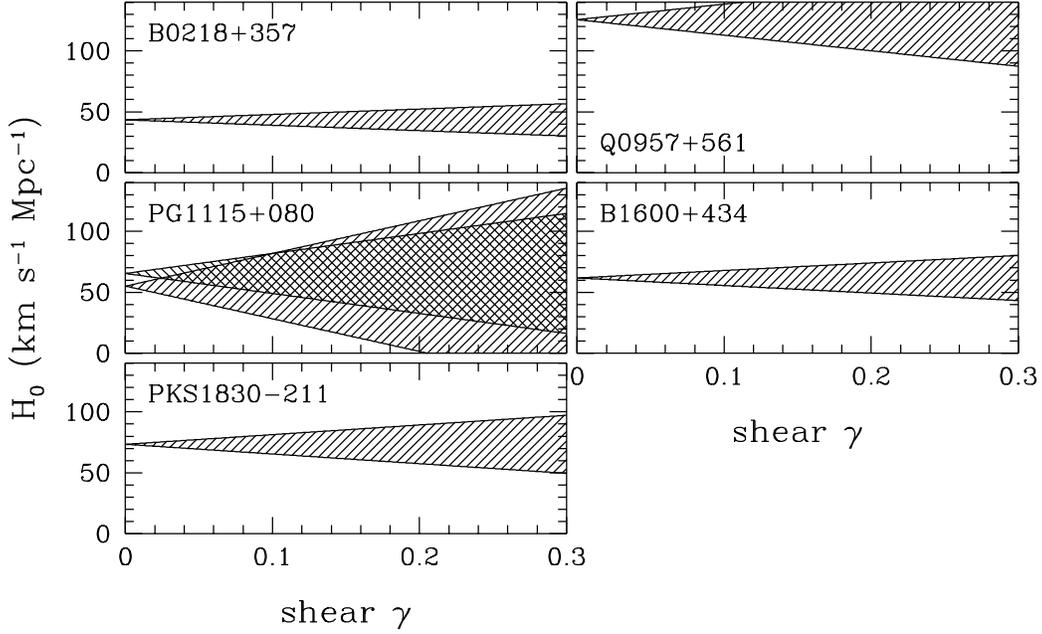}}
\caption{
Effects of external shear on $\H0$ estimates.  Each panel shows
the bounds on $\H0$ as a function of the shear amplitude for the
specified lens.  (This figure does not include statistical
uncertainties due to observational errors.)  The bounds are
computed using eqs.~(\ref{eq:Tbounds1}) and (\ref{eq:Tbounds2}),
assuming no knowledge of the angle between the shear and the
lens galaxy major axis.  The bounds would be narrower if this
angle were constrained.  For PG~1115+080, the two sets of
bounds refer to the two different time delays.  All results are
computed for a cosmology with $\Omega_M=1$ and $\Omega_\Lambda=0$
but are nearly the same for other cosmologies (see Figure 1).
}
\end{figure}

\end{document}